\documentclass[11pt]{iopart}
\usepackage{iopams} 
\usepackage{graphicx}
\usepackage[colorlinks=true,linkcolor=blue,citecolor=red,urlcolor=magenta]{hyperref}
\usepackage[dvipsnames]{xcolor}
\usepackage{charter} 
\usepackage[top=3cm,bottom=3cm,left=2cm,right=2cm,a4paper]{geometry}

\begin{document}  

\title[Dynamical resistivity of a few interacting fermions]{Dynamical resistivity of a few interacting fermions \\ to the time-dependent potential barrier}

\author{Dillip K. Nandy \& Tomasz Sowi\'{n}ski}

\address{Institute of Physics, Polish Academy of Sciences \\ Aleja Lotnik\'ow 32/46, PL-02668 Warsaw, Poland}
  
\begin{abstract}
We study the dynamical response of a harmonically trapped two-component few-fermion mixture to the external gaussian potential barrier moving across the system. The simultaneous role played by inter-particle interactions, rapidity of the barrier, and the fermionic statistics is explored for systems containing up to four particles. The response is quantified in terms of the temporal fidelity of the time-evolved state and the amount of quantum correlations between components being dynamically generated. Results are also supported by analysis of the single-particle densities and temporal number of occupied many-body eigenstates. In this way, we show that the dynamical properties of the system crucially depend on non-trivial mutual relations between temporal many-body eigenstates, and in consequence, they lead to volatility of the dynamics. Counterintuitively, imbalanced systems manifest much higher resistivity and stability than their balanced counterparts. 
\end{abstract}

\section{Introduction}
Isolated quantum systems described by time-dependent Hamiltonians, due to explicit violation of the energy conservation, manifest various phenomena that are inaccessible by their stationary counterparts. To capture their properties accurately one needs to understand their dynamical response to the external driving. This is particularly crucial in the case of state-of-the-art atomic physics experiments where extremely precise coherent control of strongly correlated quantum systems is tremendously rigorous and is treated as the specific dream come true \cite{1993WarrenScience}.

One of the typical directions considered for coherent manipulation of atomic systems is related to periodic driving \cite{2017EckardtRMP}. On different levels, it has been demonstrated that such forcing can be used as a powerful tool for precise controlling of quantum dynamics and it gave a ride to a plethora of fundamentally important and beautiful results. To mention only several examples: the dynamical control of the superfluid-Mott insulator transition \cite{Zenesini}, the AC-induced coherent resonant tunneling \cite{Ivanov, Sias}, expansion dynamics of Bose-Einstein condensates through periodic driving \cite{Lignier}, creation of artificial gauge fields \cite{Aidelsburger1, Aidelsburger2}, the realization of topological insulators via time-periodic potentials \cite{Jotzu, Aidelsburger3}, the coherent coupling between different energy bands \cite{Gemelke, Ha}, or frustrated magnetism \cite{2011StruckScience}. On the theoretical side, the underlying concept which gives an adequate description of a periodically driven quantum system is the Floquet theory \cite{Shirley, Sambe, Grifoni} which gives a nice tool for an appropriate description of the system stroboscopically, {\it i.e.}, after each period of the driving. 

External driving is a commonly used concept in atomic physics problems (see \cite{2017EckardtRMP,2011PolkovnikovRMP} for reviews) and recently it was also exploited in the limit of a small but well-determined number of particles \cite{2016VolosnievPRB,2016EbertAnnPhys,2016YinPRA,2016GharashiPRA,2020MukherjeePRA}. However, still, it is not fully clear how different correlations of such systems are dynamically generated during such a process. One of the natural questions is on collectiveness of different properties, {\it i.e.}, how different few-body results start to resemble their many-body counterparts when the number of particles is successively increased. In our work, we make another step in this direction and we focus on transmutations of the system's properties enforced during a single period of external driving. First, we start with an interacting ground state of a system of a few interacting fermions confined in an external static trap. Then the system is non-adiabatically affected by a moving external potential barrier that crosses the system back and forth between the edges of the trap. In this way we investigate how the system dynamically transits between the two well-explored stationary cases, {\it i.e.}, confined in a pure harmonic oscillator \cite{2013AstrakharchikPRA,2013SowinskiPRA,2013GharashiPRL,2014LindgrenNJP,2014DeuretzbacherPRA,2014HarshmanPRA,2014DAmicoJPhysB} and in a double-well \cite{2016SowinskiEPL,2017TylutkiPRA,2017HarshmanPRA,2018ErdmannPRA,2018BurchiantiPRL,2019ErdmannPRA}. Depending on the rapidity of the barrier we observe how the state of the system is pulled away from the temporal ground state. By quantifying this distortion with a well-defined temporal fidelity, we capture the non-trivial interplay between external forcing, quantum statistics, number of particles, and interaction strength. The sweeping barrier protocol proposed here can be viewed as a few-body counterpart of the scenario recently realized experimentally in the many-particle limit with $^6$Li atoms \cite{2020KwonScience}. It gave a route to the beautiful observation of the DC Josephson supercurrent as previously predicted theoretically \cite{2019ZaccantiPRA}.

Since our purely theoretical work is motivated by seminal experiments with two-component mixtures of few fermionic $^6$Li atoms confined in quasi-one-dimensional external traps \cite{Serwane, Zurn, Wenz} we focus on equal-mass components. However, generalization to different-mass atoms (as also conceivable experimentally \cite{2010TieckePRL,2011NaikEPJD,2015CetinaPRL,cetina2016ultrafast,Grimm2018DyK}) is straightforward. Since perfect control and manipulation of these systems were convincingly demonstrated we believe that our findings may have importance for further experimental and theoretical progress (for general reviews see \cite{2016OnofrioReview,2016ZinnerRev,2019SowinskiRPP}).

\section{The Model}
In the present study we focus on a few two-component ultra-cold fermionic atoms that interact through short-range delta-like potential initially confined in a harmonic trap. In addition to the harmonic potential we introduce a gaussian potential barrier which periodically moves across the system. We aim to study a resistivity of the system properties to this disturbance depending on its rapidity. This specific situation can be modeled by the following many-body Hamiltonian
\begin{eqnarray} \label{Hamiltonian}
 \hat{\mathcal{H}}(x_0) = \sum_{\sigma} \int\!\!\mathrm{d}x\,\hat{\Psi}^{\dagger}_{\sigma}(x) \left[H_{0} + U(x_0)\right] \hat{\Psi}_{\sigma}(x)
+ g \int \mathrm{d}x\,\hat{\Psi}^{\dagger}_{\uparrow}(x) \hat{\Psi}^{\dagger}_{\downarrow}(x) \hat{\Psi}_{\downarrow}(x)  
 \hat{\Psi}_{\uparrow}(x),
\end{eqnarray}
where the single-particle part of the Hamiltonian contains two parts $H_0$ and $U(x_0)$. The first represents the static harmonic oscillator part of the standard form
\begin{equation}
H_0 = -\frac{\hbar^2}{2m}\frac{\mathrm{d}^2}{\mathrm{d}x^2}+\frac{m \Omega^2}{2}x^2.
\end{equation}
The second describes the additional gaussian barrier of the width $\beta$ and height $U_0$ centered around (in principle time-dependent) position $x_0$. It can be written as
\begin{equation}
U(x_0)= \frac{U_0}{\sqrt{\pi}\beta} \mathrm{e}^{-(x-x_0)^2/\beta^2}.
\end{equation}
From now, we will use the natural units of the harmonic oscillator which means that energies, lengths, and momenta have units of $\hbar\Omega$, $\sqrt{\hbar/m\Omega}$, and $\sqrt{\hbar m \Omega}$, respectively. In our work we focus on the case of relatively narrow and quite high barrier, {\it i.e.}, in these units we set $U_0=4$ and $\beta=0.2$, respectively. However, generalization to other barriers is straightforward.

In the Hamiltonian (\ref{Hamiltonian}), the fermionic field operators $\hat{\Psi}_{\sigma}(x)$ annihilate fermion of component $\sigma\in\{\downarrow,\uparrow\}$ at position $x$, while $g$ is an effective interaction coupling between fermions belonging to opposite components. The field operators obey standard anti-commutation relations $\{\hat{\Psi}_\sigma(x),\hat{\Psi}_{\sigma'}^\dagger(x')\}=\delta_{\sigma\sigma'}\delta(x-x')$ and $\{\hat{\Psi}_\sigma(x),\hat{\Psi}_{\sigma'}(x')\}=0$. It is clear that independently on the position of the barrier $x_0$ and its motion the Hamiltonian (\ref{Hamiltonian}) commutes with operators of the number of particles in individual components, $\hat N_\sigma=\int\!\mathrm{d}x\,\hat{\Psi}^{\dagger}_{\sigma}(x)\hat{\Psi}_{\sigma}(x)$. Therefore in the following, we analyze properties of the system in the eigensubspaces of fixed $\hat{N}_\downarrow$ and $\hat{N}_\uparrow$.  

In the following, a whole analysis will be carried out numerically in the static, barrier-independent single-particle basis $\{\phi_i(x)\}$ being the eigenbasis of the pure harmonic oscillator Hamiltonian $H_0$. Therefore, after decomposing field operators in this basis as following
\begin{eqnarray}
 \hat{\Psi}_{\downarrow}(x) = \sum_i \phi_{i}(x) \hat{a}_{i}, \ \
 \hat{\Psi}_{\uparrow}(x) = \sum_i \phi_{i}(x)\hat{b}_{i}, 
\end{eqnarray}
where $\hat{a}_i$ and $\hat{b}_i$ are annihilation operators for the $\downarrow$ and $\uparrow$ fermions at state $\phi_i(x)$, one rewrites the many-body Hamiltonian (\ref{Hamiltonian}) to the form
\begin{eqnarray} \label{Hamiltonian2}
\hat{\mathcal{H}}(x_0) &= \sum_i \left(i+\frac{1}{2}\right)\left(\hat{a}^{\dagger}_{i}\hat{a}_{i} + \hat{b}^{\dagger}_{i}\hat{b}_{i}\right) 
+ \sum_{ij} U_{ij}(x_0) \left[\hat{a}^{\dagger}_{i}\hat{a}_{j} + \hat{b}^{\dagger}_{i}\hat{b}_{j} + h.c.\right] \nonumber \\
&+ g\sum_{ijkl} I_{ijkl} \hat{b}^{\dagger}_i\hat{a}^{\dagger}_j \hat{a}_k\hat{b}_l.
\label{Ham}
\end{eqnarray}
where $U_{ij}(x_0) = \int\!\mathrm{d}x\,\phi^*_{i}(x) U(x_0) \phi_{j}(x)$ are matrix elements of the barrier potential in the chosen single-particle basis, while the coefficients $I_{ijkl}$ are determined by the interaction part of the Hamiltonian and they are given by $I_{ijkl} = \int\!\mathrm{d}x \phi^*_i(x)\phi^*_j(x) \phi_k(x) \phi_l(x)$. Due to the mirror left-right symmetry of the harmonic oscillator eigenstates these integrals are nonzero only when corresponding sums of the indices $(i + j + k + l)$ are even. 

Since the single-particle basis is independent of the position of the barrier, the whole static (but barrier-dependent) Hamiltonian (\ref{Hamiltonian}) can be simply converted to the time-dependent one by varying the position of the barrier $x_0$ in time. We will consider the simplest scenario of a periodic motion of the barrier, $x_0(t) = A\cos(\omega t)$, and we will discuss properties of the system for different frequencies $\omega$ after one period $T=2\pi/\omega$. To make sure that initially the system is not influenced by the barrier we set $A=4$. In this case, the ground-state of the system with the barrier in its initial position is in fact the same as the ground-state of the system in a pure harmonic oscillator. Taking all these initial conditions into account, we investigate properties of the system by solving the time-dependent Schr\"odinger equation, $i\frac{d}{dt}|\psi(t) \rangle = \hat{\mathcal{H}}(x_0(t)) |\psi(t) \rangle$, with the initial condition $|\psi(t = 0) \rangle = |\mathtt{G}_0\rangle$, where $|\mathtt{G}_0\rangle$ is a ground-state of the interacting system confined in a pure harmonic oscillator. 

The computational scheme used for solving the time-dependent many-body Schr\"odinger equation is based on representing the Hamiltonian (\ref{Hamiltonian2}) as a matrix in the Hilbert space spanned by Fock states with the lowest single-particle excitations. In our case, the Fock basis is time-independent and it is built from $M$ the lowest single-particle eigenstates of a pure harmonic oscillator. For the cases studied here we set $M=20$ and $18$ for $(N_\uparrow,N_\downarrow)=(1,1)$ and $(2,1)$, respectively, and $M=15$ for $(N_\uparrow,N_\downarrow)=(3,1)$, and $(2,2)$. It means that the dimension of the Fock space in these cases is respectively $D=400, 2754, 6825$, and $11025$. The cut-off $M$ is selected in such a way that, for interactions and other parameters of the problem considered, the final results do not change noticeably when $M$ is increased albeit numerical calculations are still feasible in a reasonable computational time and with reasonable computational resources. After determining all matrix elements of the Hamiltonian in this basis (some of them are time-dependent due to the moving barrier), the time-dependent Schr\"odinger equation is solved numerically by the fourth-order Runge-Kutta method with a time step $\delta t=0.004$ in chosen units of the harmonic oscillator. We also checked that the final results do not change if the time step is decreased to $\delta t = 0.002$ and $\delta t =0.001$. To get a better understanding of the system properties we also calculate the many-body spectra of a given system. For this purpose, we use the same matrix representation of the Hamiltonian as used for time-dependent calculations. For fixed barrier positions $x_0$ we numerically diagonalize the corresponding matrix via the Implicitly Restarted Arnoldi Method \cite{ArnoldiBook}. In this way, we find the lowest eigenenergies as functions of barrier position (with the barrier's position step $\delta x_0=0.01$) and we visualize the temporal spectrum of the system.

The main aim of the present study is to understand the response of a few interacting systems to a potential barrier moving across a harmonic trap. It is clear that in the case of adiabatic motion of the barrier ($\omega\rightarrow 0$), at any moment the system remains in its temporal ground state $|\mathtt{G}(x_0(t))\rangle$, {\it i.e.}, in the many-body ground state of the Hamiltonian $\hat{\cal H}(x_0(t))$. It means that after a whole period, $T=2\pi/\omega\rightarrow \infty$, the system finally returns to the initial ground state $|\mathtt{G}_0\rangle$ and thus the final fidelity ${\cal F} = |\langle\psi(t=T)|\psi(t=0) \rangle|^2$ is exactly equal to unity. Surely, the situation is different when the barrier makes a loop in a finite period of $T$ since then the system can be partially excited to other many-body states. To capture this possibility we introduce the temporal fidelity 
\begin{equation}
{\cal F}_\omega(t) = |\langle\psi(t)|\mathtt{G}(x_0(t))\rangle|^2,
\end{equation}
which quantifies the overlap of the transient state of the system $|\psi(t)\rangle$ with the ground state $|\mathtt{G}(x_0)\rangle$ of the temporal Hamiltonian $\hat{\cal H}(x_0(t))$. In this way, we can quite precisely determine moments when the system becomes excited to higher many-body states and associate these excitations with specific properties of the Hamiltonian's spectrum. We suspect that in the adiabatic limit, $\omega\rightarrow 0$, the temporal fidelity ${\cal F}_\omega(t)\rightarrow 1$ for any moment $t$.

At this point, we emphasize that the temporal fidelity ${\cal F}_\omega(t)$ is not the quantity that can be measured directly in state-of-the-art experiments. It is rather a theoretical concept giving the most natural way to quantify the distance in the Hilbert space between different many-body states. However, after appropriate repetition of high-precision measurements, theoretical methods of the quantum state tomography \cite{2003ArianoREVIEW,2017LanyonNatPhys} very often give a route to deduce a form of the considered many-body state. Thus, in principle, the temporal fidelity can be obtained indirectly. In the case of confined systems, it can be done by measuring different probabilities of finding particles occupying different single-particle orbitals and comparing them with corresponding probabilities describing the ground state of the system. In experimental realizations with few interacting particles confined in one-dimensional traps, such measurements are currently accessible \cite{Serwane}.

\begin{figure}
\centering
\includegraphics{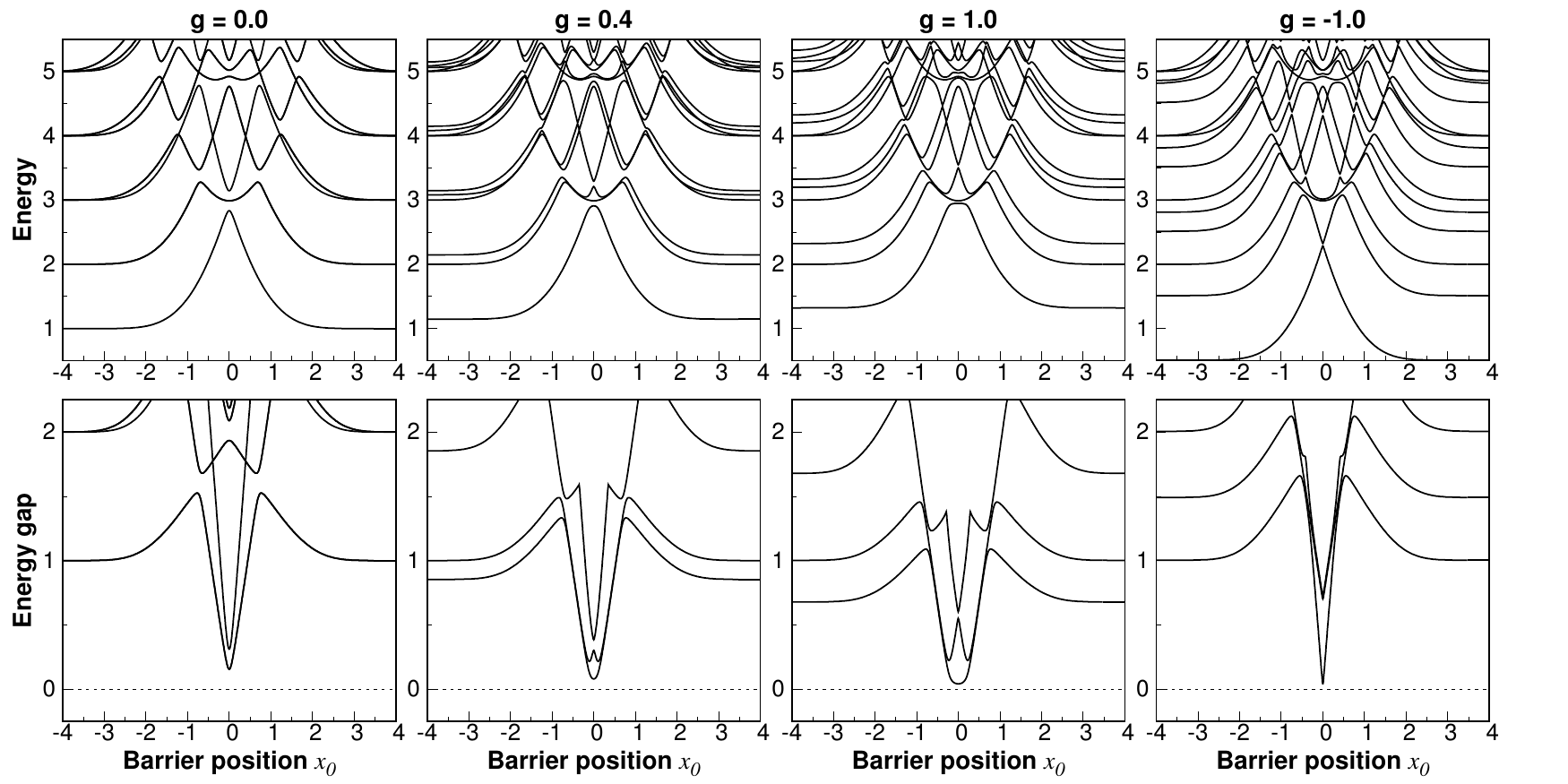}
\caption{ {\bf (top row)} The many-body spectrum of the Hamiltonian $\hat{\cal H}(x_0)$ for $N_{\uparrow} = N_{\downarrow} = 1$ particles as a function of barrier position $x_0$ and different interaction strengths. For the barrier being far away from the center ($x_0=\pm 4$) the spectrum resembles the corresponding spectrum of a pure harmonically trapped system. When the barrier is close to the center, a quasi-degeneracy between many-body states, characteristic for double-well confinement, is visible. Note, that for stronger interactions the energy gap at $x_0=0$ becomes significantly reduced. {\bf (bottom row)} Energy gaps between the lowest three excited states and the temporal ground-state of the Hamiltonian $\hat{\cal H}(x_0)$. It is clear that along with increasing interactions (repulsive as well as attractive) energy gaps closes. Note however that in the case of repulsive forces the lowest excited state, due to inappropriate exchange symmetry, is not dynamically coupled and does not contribute to the dynamics (see main text for detailed explanation). Thus, an effective gap to the lowest relevant excited state grows with increasing repulsions.} 
\label{Fig1}
\end{figure}
\section{Two-particle system}
To make our analysis as clear as possible let us start from the simplest system of $N_\uparrow=N_\downarrow=1$ particles. In Fig.~\ref{Fig1} we display the temporal spectrum of the many-body Hamiltonian $\hat{\cal H}(x_0)$ as a function of the barrier's position for different inter-particle interaction strengths. It is clear that whenever the barrier is far from the center of the trap ($x_0=\pm 4$) the spectrum resembles a well-known spectrum of two particles confined in the harmonic confinement \cite{1998BuschFoundPhys} and the ground state is well-isolated from other many-body eigenstates. When the barrier approaches the center of the trap ($x_0\approx 0$), a characteristic quasi-degeneracy of the spectrum is formed. This is a natural consequence of a quasi-degeneracy of the single-particle spectrum in a double-well potential \cite{2016SowinskiEPL,Nandy_2020}. It is important to note that along with increasing interactions, the energy gap at $x_0=0$ between the ground-state and excited many-body states becomes smaller. Since this gap has fundamental significance for excitations when a finite-time transition of the barrier is considered, one can suspect that for sufficiently strong interactions keeping adiabaticity (understood as remaining of the state of the system in the temporal ground state) may be very challenging.

\begin{figure}
\centering
\includegraphics{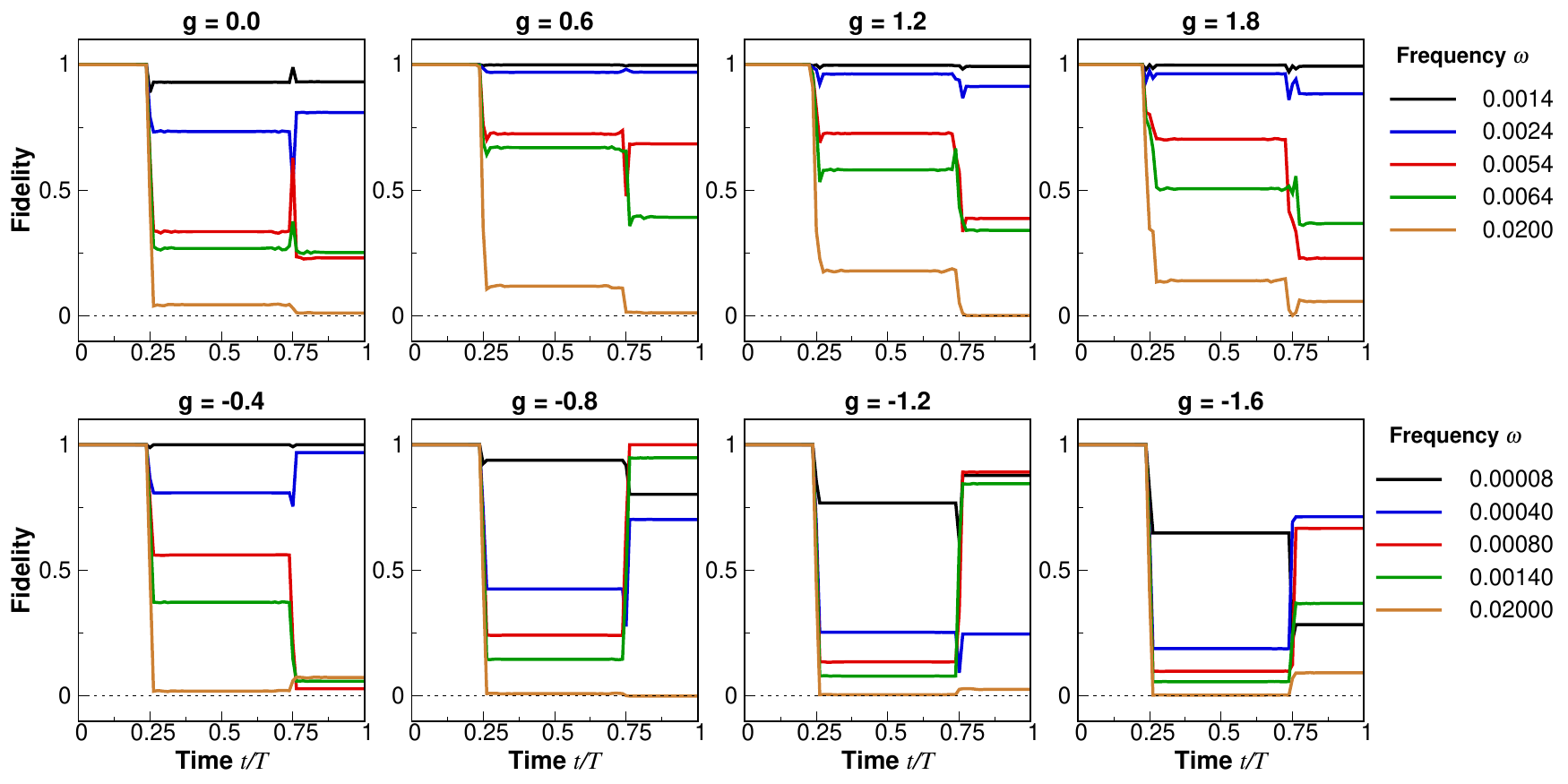}
\caption{Time evolution of the temporal fidelity ${\cal F}_\omega$ for different interactions $g$ and different frequencies $\omega$ for the system of $N_{\uparrow} = N_{\downarrow} = 1$. For better comparison, time is expressed in units of the driving period $T$. Note rapid changes of the fidelity when the barrier crosses the center of the trap. To avoid any misunderstandings, note that the same colors correspond to {\it different} frequencies for attractive and repulsive interactions. } \label{Fig2}
\end{figure}
To get a better and quantitative understanding of the dynamical response of the system depending on the barrier frequency $\omega$, we calculate the temporal fidelity ${\cal F}_\omega(t)$ at different time-frames. These results are shown in Fig.~\ref{Fig2} for some chosen values of interactions and different driving frequencies. For convenience, we compare results for different frequencies $\omega$ by plotting fidelities as functions of time $t$ rescaled with respect to the driving period $T$. It is clear that in the first stage of the dynamics, {\it i.e.}, before the barrier transits through the area occupied by particles, the system remains unaffected by the barrier position. Therefore, in the initial stage, the system is completely described by the ground state of the time-independent Hamiltonian with pure harmonic oscillator and thus the fidelity is close to unity irrespective of the sign of interactions and the frequency $\omega$. Then, when the barrier is in the vicinity of the center of the trap, the fidelity rapidly decreases for both, repulsive and attractive, interactions. Of course in the adiabatic limit (very small frequency $\omega$) the fidelity does not change significantly signaling that the system remains in the temporal ground state. Note however that the magnitude of a fidelity's drop depends on the sign of interactions and it is much deeper in the attractive case. This observation can be directly related to the properties of the many-body spectrum. As clearly seen in Fig.~\ref{Fig1}, the energy gap between the many-body ground state and other excited states closes much differently in the case of attractive interactions and it is more sensitive to the strength $g$. Moreover, it turns out that for repulsive interactions the lowest excited state does not contribute to the dynamics of the system. This is directly related to the fact that this excited state, in contrast to the initial state $|\mathtt{G}_0\rangle$, is anti-symmetric with respect to the exchange of particles' position. Since the Hamiltonian of the balanced system of two particles commutes with this particular symmetry, the states having different symmetry properties are not coupled during the evolution independently on the frequency of the driving. Consequently, an effective gap to the first contributing many-body state is essentially larger. Therefore, to get a significant fidelity drop caused by the first transition of the barrier, one needs to consider scenarios being much further from the adiabatic limit (larger frequencies $\omega$). The situation is reversed in the case of attractive forces. In this case, the first excited state is symmetric under the exchange of particles. Therefore, if the energy gap is small enough, it is easily coupled to the temporal state of the system. It means that the repulsively interacting system is much more robust to the driving than its attractively interacting counterpart.
\begin{figure}
\centering
\includegraphics{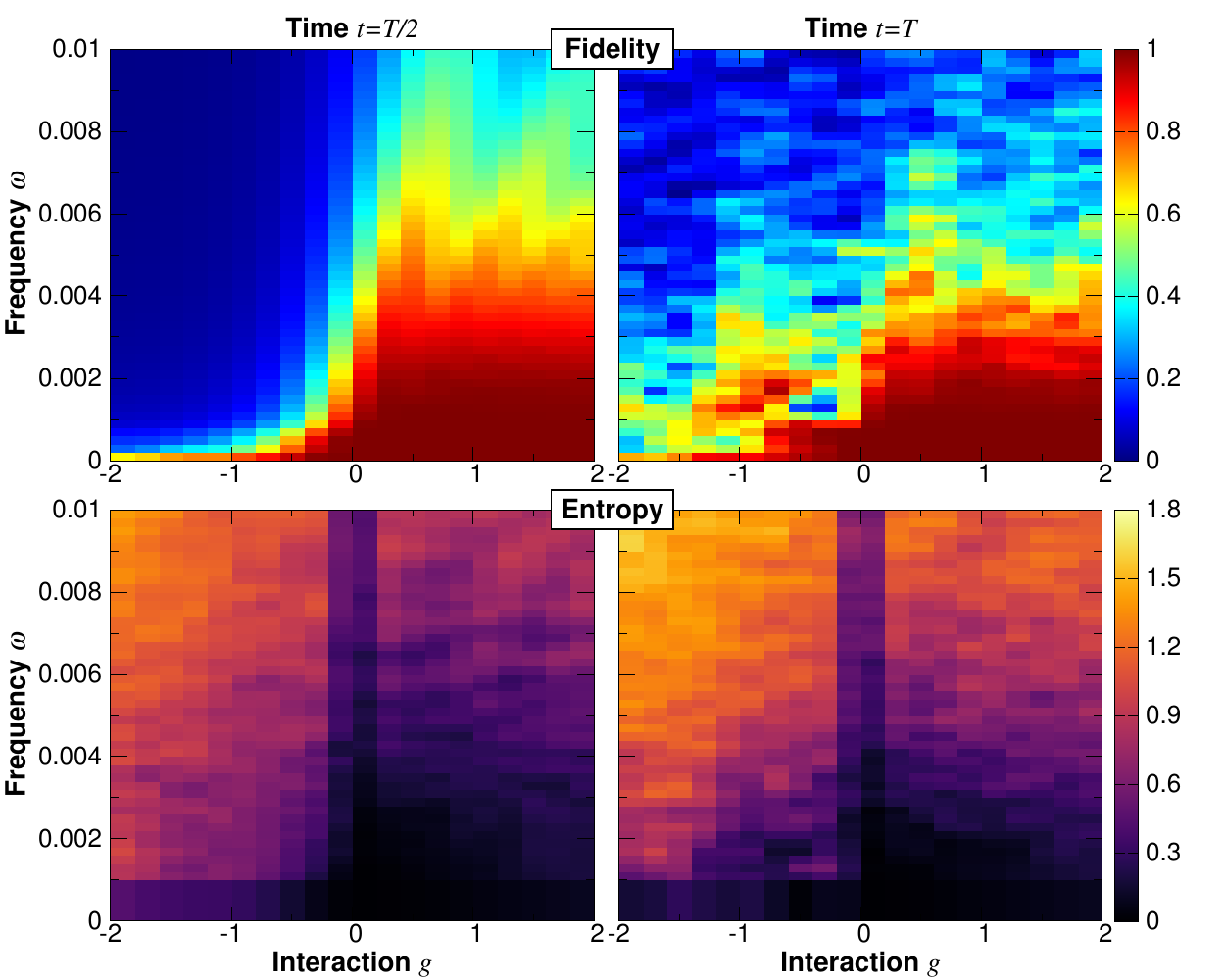}
\caption{The temporal fidelity ${\cal F}_\omega$ (top row) and the inter-component entanglement entropy $S$ (bottom row) calculated for the system of $N_\uparrow=N_\downarrow=1$ particles at time $t=T/2$ (left column) and $t=T$ (right column) as a function of interaction strength and driving frequency $\omega$. Chaotic behavior of the fidelity at the final instant $t=T$ is clearly visible. It indicates unpredictability of the dynamics after the second transition of the barrier through the trap. In contrast, after the first transition ($t=T/2$) the dynamics is much more regular and predictable. In both cases, departure from the adiabatic limit is much more intensive for attractive forces. } 
\label{Fig3}
\end{figure}

After the barrier fully transits through the system we observe again that the fidelity does not change in time until returning transition. This occurs for all frequencies, irrespectively on the sign of interactions. Again, this behavior can be explained by a closer inspection of the many-body spectrum. After the transition, decomposition amplitudes of the temporal state of the system to the temporal many-body eigenstates almost do not change since gaps between individual states are much larger than the energy scale defined by the frequency $\omega$. In consequence, an overlap of the temporal state with the temporal ground state of the system remains almost constant. 

In Fig. \ref{Fig3}, we present snapshots of the temporal fidelity at $T/2$ and $T$ as a function of driving frequency and the interaction strength. These plots compare the response of the system to single and double transition of the moving barrier. Of course, in both cases, in the adiabatic limit ($\omega\rightarrow 0$) the fidelity remains equal to $1$. Nevertheless, for attractive interactions achieving adiabaticity requires much smaller frequencies. As can be seen, after the first transition of the barrier ($t=T/2$), the fidelity is rather a smooth function of interactions and frequency. In contrast, at the final moment ($t=T$) the dependence on these parameters is highly non-trivial and in practice unpredictable. Random behavior of the fidelity at time $t=T$ is related to non-obvious relative dynamics between the internal motion of the system and the motion of the barrier. Namely, after the first transition of the barrier, the system evolves almost independently on the barrier's position. Depending on frequency $\omega$ the second transition effectively happens at quite randomly chosen moment of this evolution. This is particularly the case when after the first transition the system becomes essentially excited (far from the adiabatic limit) and its evolution is highly non-trivial. 

It is very instructive to notice that the situation is substantially different when, instead of the fidelity, one considers quantum correlations induced by the moving barrier. For example, let us focus on the inter-component entanglement entropy defined as
\begin{equation}
S(t) = -\mathrm{Tr}_\uparrow \left[\hat\rho_\downarrow(t) \ln\hat \rho_\downarrow(t)\right]=-\mathrm{Tr}_\downarrow \left[\hat\rho_\uparrow(t) \ln\hat \rho_\uparrow(t)\right],
\end{equation}
where the temporal reduced density matrix of a chosen component $\hat{\rho}_\sigma(t)$ can be calculated straightforwardly by tracing out remaining component $\sigma'$ as $\hat\rho_\sigma(t)=\mathrm{Tr}_{\sigma'} \left[|\Phi(t)\rangle\langle\Phi(t)|\right]$. Analogously as we did for the fidelity, in the bottom row of Fig.~\ref{Fig3} we display the entropy $S(t)$ as a function of the interaction strengths and driving frequencies at two different moments $t=T/2$ and $t=T$. It is clear that for both cases the inter-component correlations are built in the system quite smoothly and nearly symmetrically on both sides of the non-interacting limit. For increasing frequencies, {\it i.e.}, when the system rides away from the adiabatic limit, correlations are almost monotonically enhanced. This regularity of the entanglement entropy can be viewed as a direct consequence of the fact that inter-component quantum correlations are related to period and strength of interactions rather than to a particular decomposition and specific evolution of the quantum state.

 It is clear that a direct experimental access to the entanglement entropy, although in principle possible (for particular examples see \cite{2012AbaninPRL,2015IslamNature}), is not straightforward. In the case of the inter-component entanglement considered here the situation is slightly simpler since experimental filtering out of a selected component can be viewed as the counterpart of a theoretical concept of tracing out of the remaining component, {\it i.e.}, the first step of the analysis. Then, by precise determination of occupation probabilities (similarly as in the case of the ground-state fidelity) and by measuring inter-particle correlations between positions of particles (possible due to very precise single-site resolution microscopes) one can try to reconstruct a reduced density matrix. In this way, entanglement entropy between components may be estimated. Nonetheless, still this is not the quantity that can be easily measured.
\begin{figure}
\centering
\includegraphics[width =\linewidth]{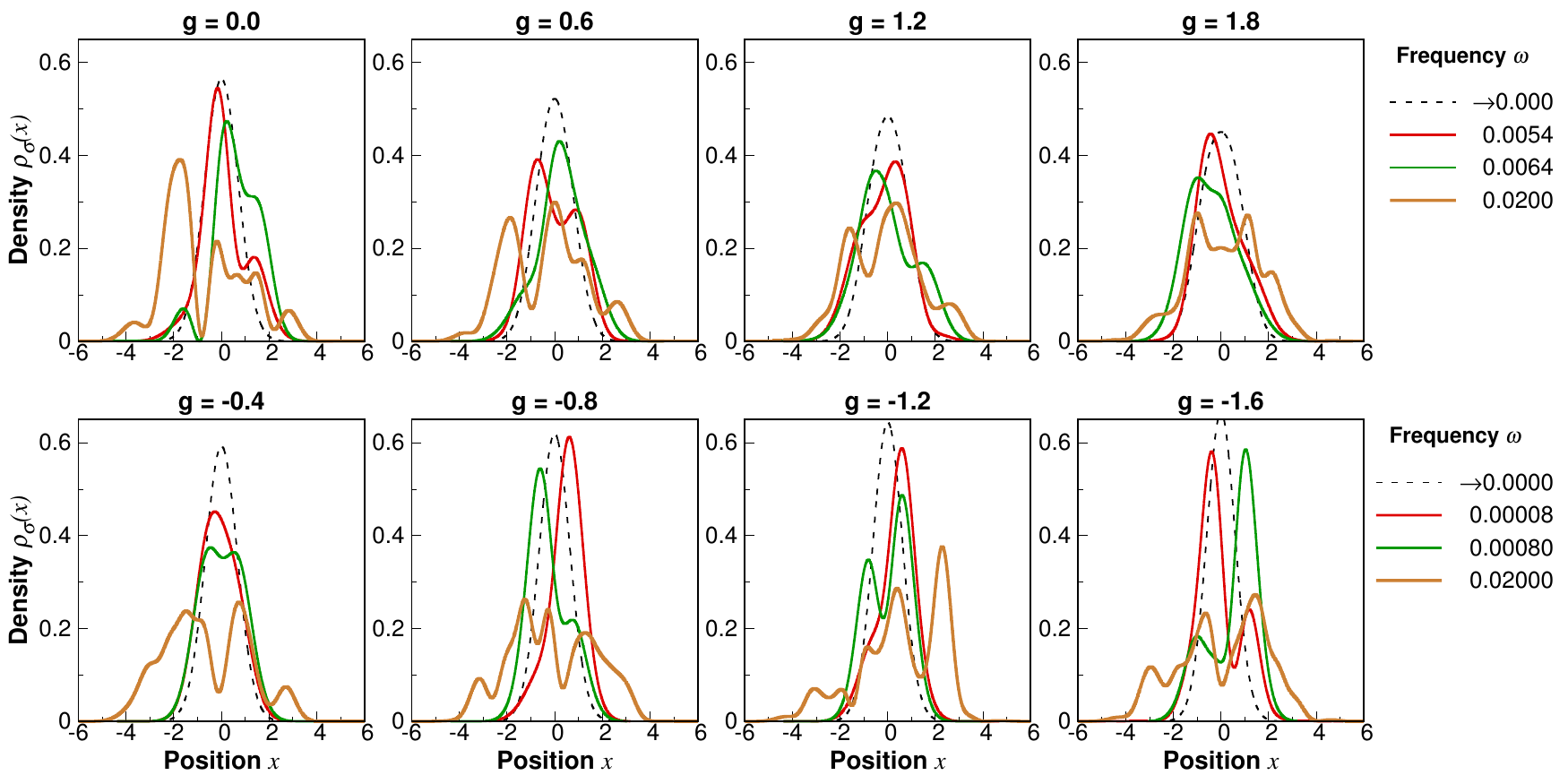}
\caption{ The single-particle density profile $\rho_\sigma(x)$ for the system of $N_{\uparrow} = N_{\downarrow} = 1$ particles calculated after the first transition of the barrier, $t=T/2$. Different colours correspond to different driving frequencies $\omega$ (compare with Fig.~\ref{Fig2}). For comparison, the density profile calculated for the many-body ground-state (adiabatic limit $\omega\rightarrow 0$) is plotted with dashed line. Along with increasing frequency and/or interactions, the system's density profile departures from the ground-state profile signaling decreasing fidelity. } 
\label{Fig4}
\end{figure}
To overcome this experimental difficulty with measuring inter-component entanglement entropy, one can consider the temporal single-particle density profile defined as $\rho_\sigma(x,t) = \langle \psi(t)|\hat\Psi^\dagger_\sigma(x)\hat\Psi_\sigma(x)|\psi(t)\rangle$. This is one of the simplest and directly measurable quantities capturing substantial deviation of the system's state $|\psi(t)\rangle$ from the temporal ground state $|\mathtt{G}(x_0(t))\rangle$. In the case of balanced system it is exactly the same for both components. In Fig.~\ref{Fig4} we plot these densities for the instant $t=T/2$ (after the first transition) and the same parameters of the driving as in Fig.~\ref{Fig2}. It is clear that whenever the fidelity becomes closer to $0$ (stronger interactions and/or larger frequencies), the density profile $\rho_\sigma(x,t)$ becomes significantly different from corresponding density obtained for a temporal ground state (black dashed line). 

\section{Spin-balanced system of four particles}
\begin{figure}
\centering
\includegraphics[width =\linewidth]{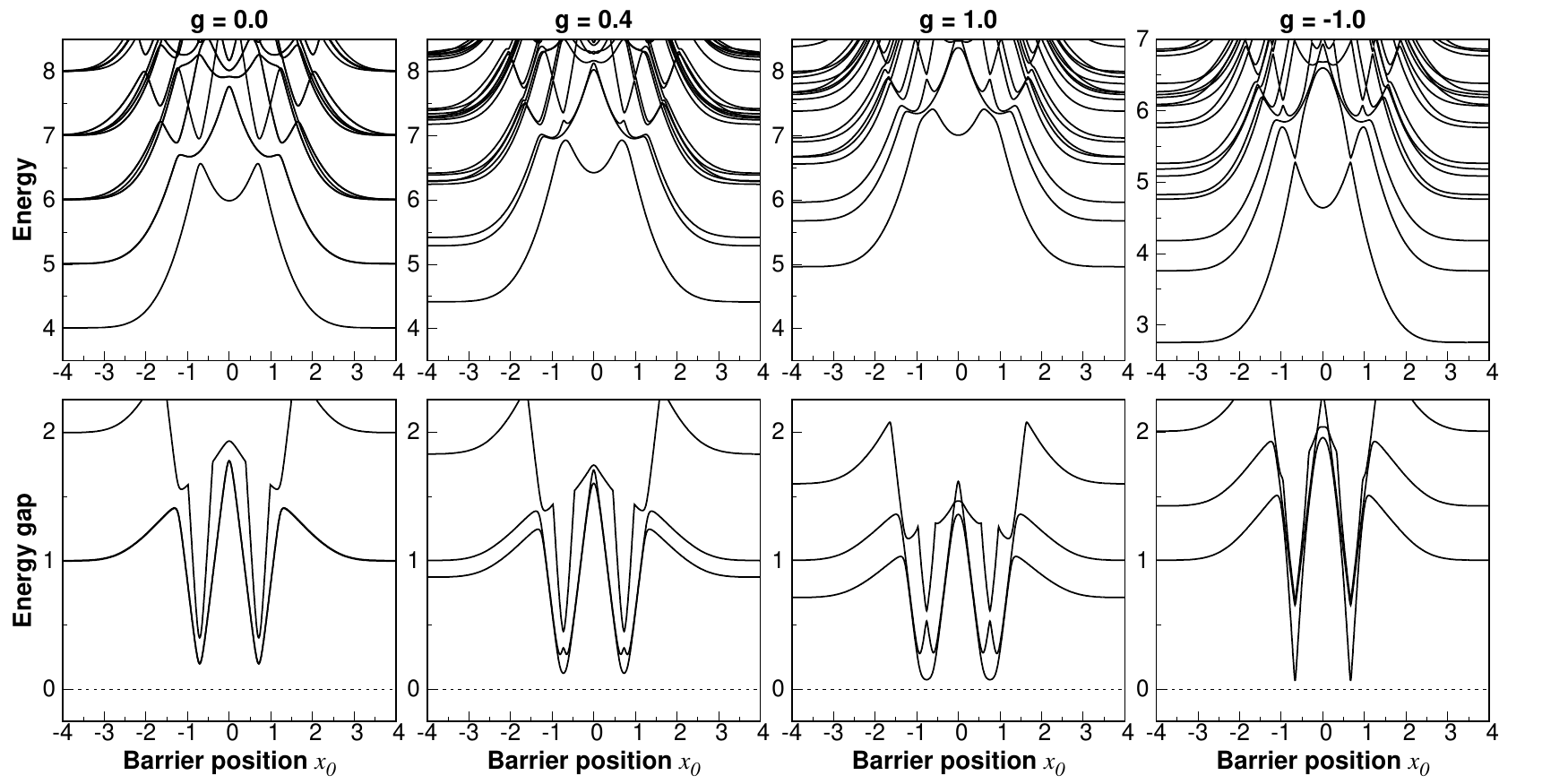}
\caption{ {\bf (top row)} The many-body spectrum of the Hamiltonian $\hat{\cal H}(x_0)$ for $N_{\uparrow} = N_{\downarrow} = 2$ particles as a function of barrier position $x_0$ and different interaction strengths. Note the substantial difference when compared to the system of two particles in Fig.~\ref{Fig1}. Here, the spectrum becomes quasi-degenerated for two distant positions of the barrier. The difference significantly changes the dynamical properties of the system. See the main text for details. {\bf (bottom row)} Energy gaps between the lowest three excited states and the temporal ground-state of the Hamiltonian $\hat{\cal H}(x_0)$. Again, when interactions are increased (repulsive as well as attractive) energy gaps become smaller.} 
\label{Fig5}
\end{figure}

\begin{figure}
\centering
\includegraphics{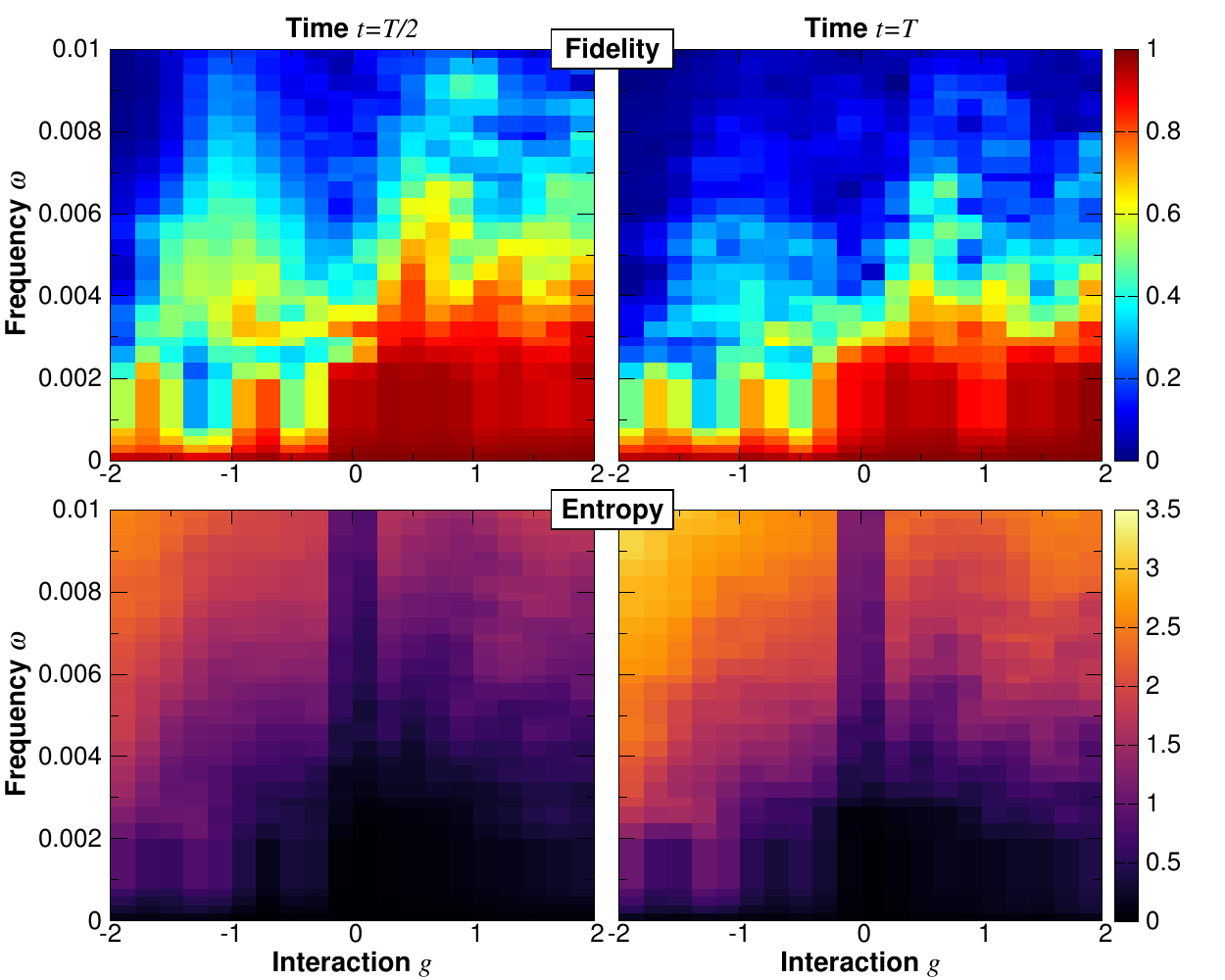}
\caption{The temporal fidelity ${\cal F}_\omega$ (top row) and the inter-component entanglement entropy $S$ (bottom row) calculated for the system of $N_\uparrow=N_\downarrow=2$ particles at time $t=T/2$ (left column) and $t=T$ (right column) as a function of interaction strength and driving frequency $\omega$. In contrast to the two particle case (Fig.~\ref{Fig3}) chaotic behavior of the fidelity is present at the final instant ($t=T$) as well as after the first transition of the barrier ($t=T/2$).} 
\label{Fig6}
\end{figure}

\begin{figure}
\centering
\includegraphics{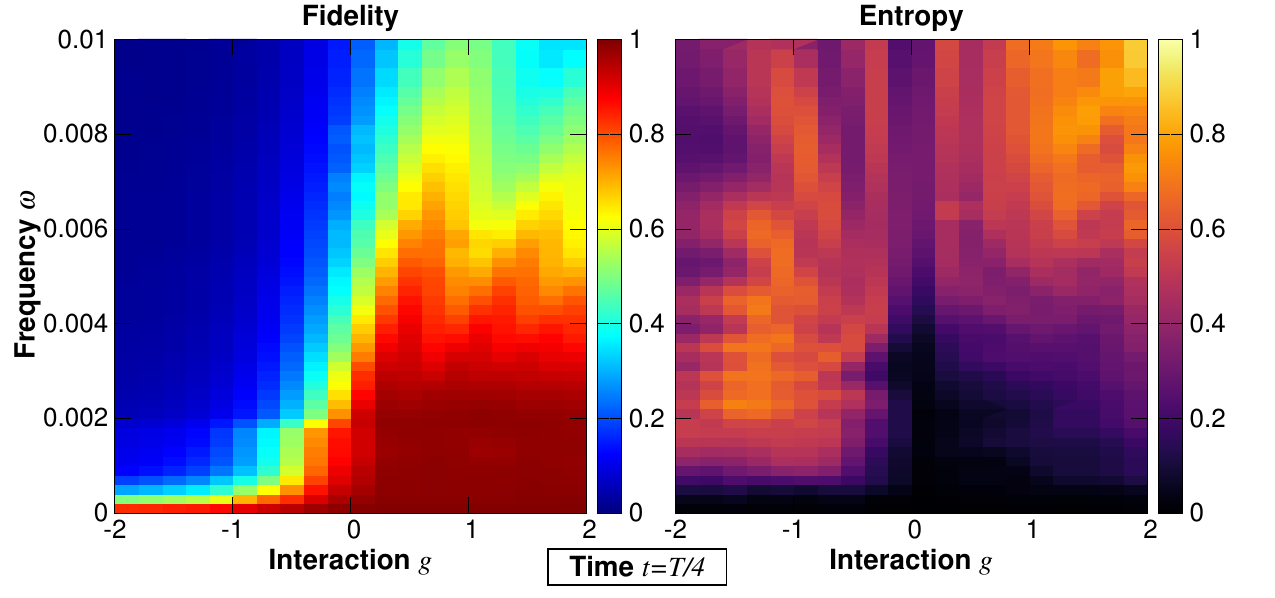}
\caption{The temporal fidelity ${\cal F}_\omega$ and the inter-component entanglement entropy $S$ calculated for the system of $N_\uparrow=N_\downarrow=2$ particles when the barrier is at the center of the trap ($t=T/4$). As explained in the main text, in this case, the quasi-degeneracy of the spectrum is reached only once and therefore the dynamical response of the system is much more regular.} 
\label{Fig7}
\end{figure}

After detailed studies of the simplest case of two distinguishable particles, it is worth considering slightly larger systems and check roles played by the quantum statistics and particle imbalance. To make it systematic, let us first focus on the balanced system with $N_\uparrow=N_\downarrow=2$. A corresponding many-body spectrum of the Hamiltonian as a function of the barrier's position and different interactions is presented in Fig.~\ref{Fig5}. It is clear that the spectrum has substantially different properties than the corresponding spectrum for a two-particle system (Fig.~\ref{Fig1}). The most prominent difference is related to the existence of two barrier positions at which the quasi-degeneracy of the many-body states is present. It may suggest that the dynamical behavior of the system may be significantly different already after the first transition of the barrier. Indeed, as clearly seen in Fig.~\ref{Fig6}, the temporal fidelity is unpredictable not only at the final moment $t=T$ but also after the first transition $t=T/2$. The phenomenological explanation of this behavior is exactly the same as previously. After the stage the first quasi-degeneracy of the many-body spectrum is present, the dynamics becomes almost independent of the barrier position. However, it remains significantly affected by particular decomposition (which in fact depends on frequency $\omega$) to different excited states. Therefore, depending on the frequency $\omega$, a moment when the stage of the second quasi-degeneracy of the spectrum is activated is highly unpredictable and leads to a very irregular pattern of the fidelity. To make sure that this general picture is correct and argumentation is valid we also check the results for the moment just after the first quasi-degeneracy is deactivated, {\it i.e.}, when the barrier is exactly at the center ($x_0=0, t=T/4$). Corresponding results are displayed in Fig.~\ref{Fig7}. It is clear, that at this instant the temporal fidelity is a very smooth function of frequency and interactions. It resembles corresponding results for $N_\uparrow=N_\downarrow=1$ at $t=T/2$ (see Fig.~\ref{Fig3}). For completeness and better comparison, in Fig.~\ref{Fig6} and Fig.~\ref{Fig7}, we also present entanglement entropy $S(t)$. In this case, similarly as in the case of two particles, quantum correlations are built in the system much more regularly and depend mostly on interaction strength and interaction period. Additionally, we checked that the response of the single-particle density profiles $\rho_\sigma(x,t)$ behaves analogously to the simpler case of two particles, {\it i.e.}, higher deviations from the temporal ground state of the system are signaled by larger disturbances of the densities.
 
\section{Spin-imbalanced systems}
\begin{figure}
\centering
\includegraphics{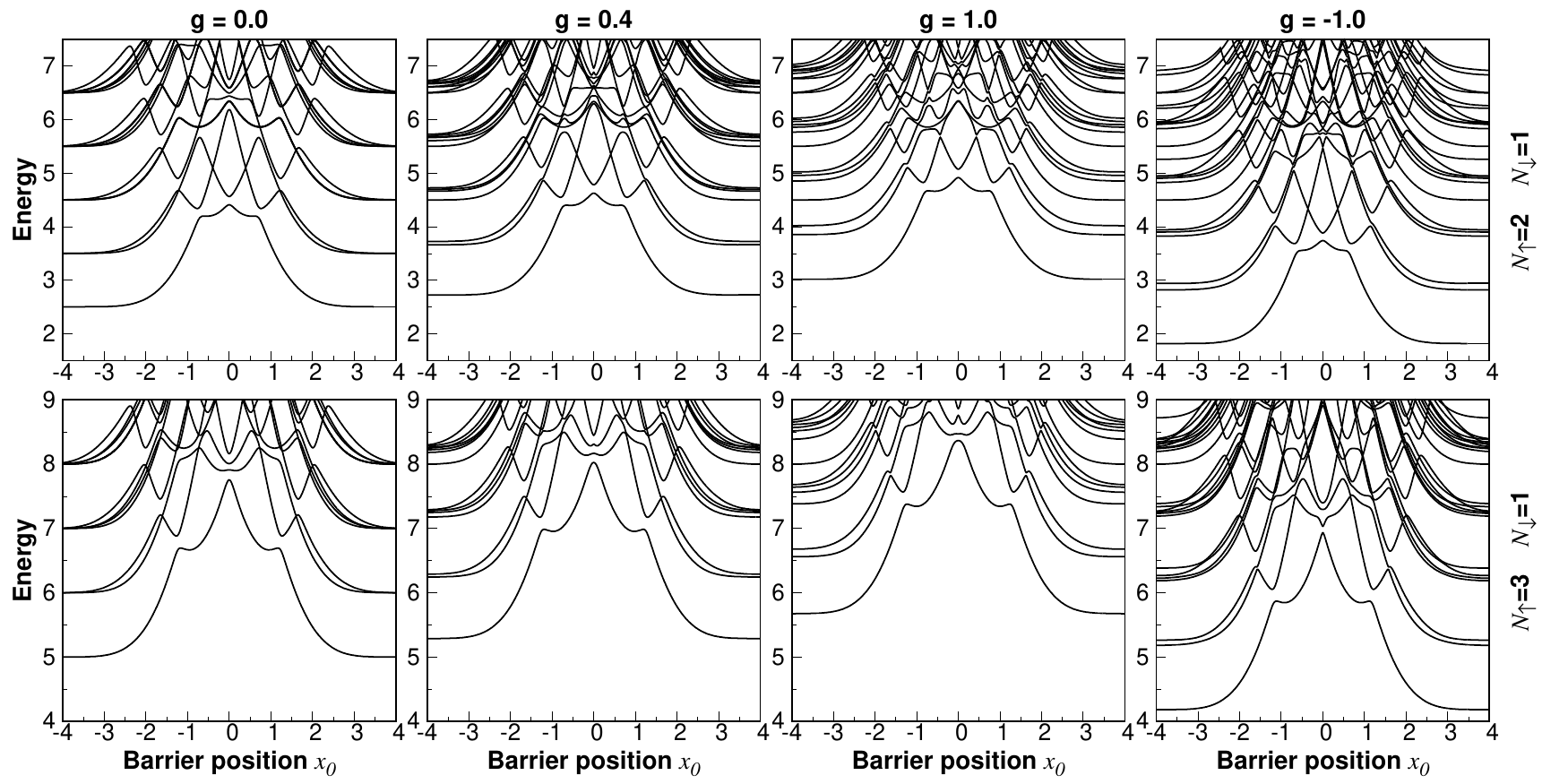}
\caption{The many-body spectrum of the Hamiltonian $\hat{\cal H}(x_0)$ as a function of barrier position $x_0$for imbalanced systems with $N_{\downarrow} = 1$ and $N_{\uparrow} = 2$ or $3$ particles and different interaction strengths. Note the qualitative difference of these spectra when compared to balanced cases (Fig.~\ref{Fig1} and Fig.~\ref{Fig5}). Here, independently on the interaction strength, the many-body ground-state is always well-isolated from excited states. This causes much better resistivity of the system to the external driving (see Fig.~\ref{Fig9}).}
\label{Fig8}
\end{figure}
\begin{figure}
\centering
\includegraphics{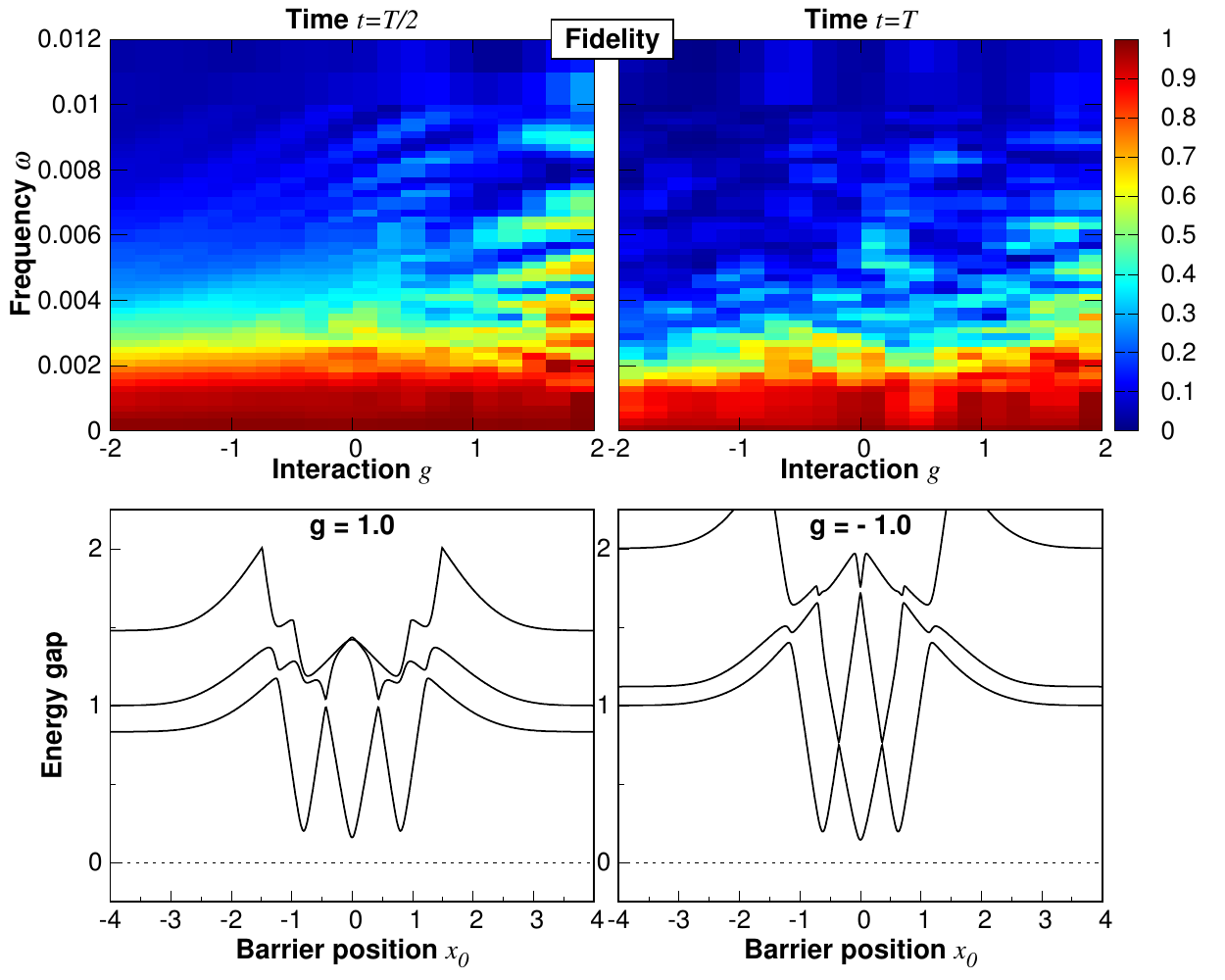}
\caption{{\bf (top row)} The temporal fidelity ${\cal F}_\omega$ for the imbalanced system of $N_\uparrow=2$ and $N_\downarrow=1$ particles at time $t=T/2$ and $t=T$ (left and right column, respectively) as a function of interaction strength and driving frequency $\omega$. It is clear that in both cases, independently of interactions, the adiabatic limit is easily reached for not too high frequencies. It means that the dynamical resistivity of the system is enhanced. {\bf (bottom row)} Energy gaps between the lowest three excited states and the temporal ground-state of the Hamiltonian $\hat{\cal H}(x_0)$ for two chosen interaction strengths in the system of $N_\uparrow=2$ and $N_\downarrow=1$ particles. In contrast to balanced systems the ground state is well-isolated also for attractive forces. This leads to mentioned resistivity of the system.} 
\label{Fig9}
\end{figure} 

Finally, let us discuss the dynamical response of the system having components with an unequal number of particles. These kinds of systems may have substantially different properties since, by the construction, they break the symmetry between components. The difference is visible already at the initial moment -- individual components have different density distributions and different spatial sizes. It means that during the evolution they become affected by the barrier at different moments and with different intensities. It should be noted, however, that although the smaller component becomes directly influenced by the barrier at later instants, it is indirectly affected earlier due to mediating interactions with the larger component being already poked by the barrier. Thus naively one can suspect that the dynamics of imbalanced systems, as being influenced by a larger number of different parameters, should be more unpredictable. In fact, this phenomenological reasoning is not fully justified. To make it clear let us first analyze the many-body spectra of these systems and their dependence on interactions and the barrier's position. We show them in Fig.~\ref{Fig8} for two systems with a total number of three and four particles. Although their general appearance is analogous to the eigenspectra of the balanced counterparts, one should note a fundamental qualitative difference -- independently of the interaction strength and the barrier position the many-body ground state is always well-isolated from the excited ones. This finite gap leads directly to  relatively high resistivity of the initial state $|\mathtt{G}_0\rangle$ to the driving, provided that the frequency $\omega$ is not too high. The system remains willingly in its temporal ground state $|\mathtt{G}(x_0(t))\rangle$ during the evolution and thus reaching the adiabatic limit is much easier. This effect is well-captured by the temporal fidelity which remains close to unity for a finite range of small frequencies independently on interactions (see Fig.~\ref{Fig9} for the system with $N_\uparrow=2$ and $N_\downarrow=1$ particles). Of course, for high enough frequencies the system becomes excited to other many-body eigenstates and the fidelity rapidly drops to 0. This drop is however rather smooth and only little irregular deviations (mainly for relatively strong interactions) are visible. Counterintuitively, it means that the imbalanced systems subjected to the external driving are much more resistive and predictable than their balanced counterparts. We checked that this fact is also well-reflected in shapes of the single-particle density profiles $\rho_\sigma(x,t)$ which for small enough frequencies resemble density profiles obtained for the temporal ground state.

\begin{figure}
\centering
\includegraphics{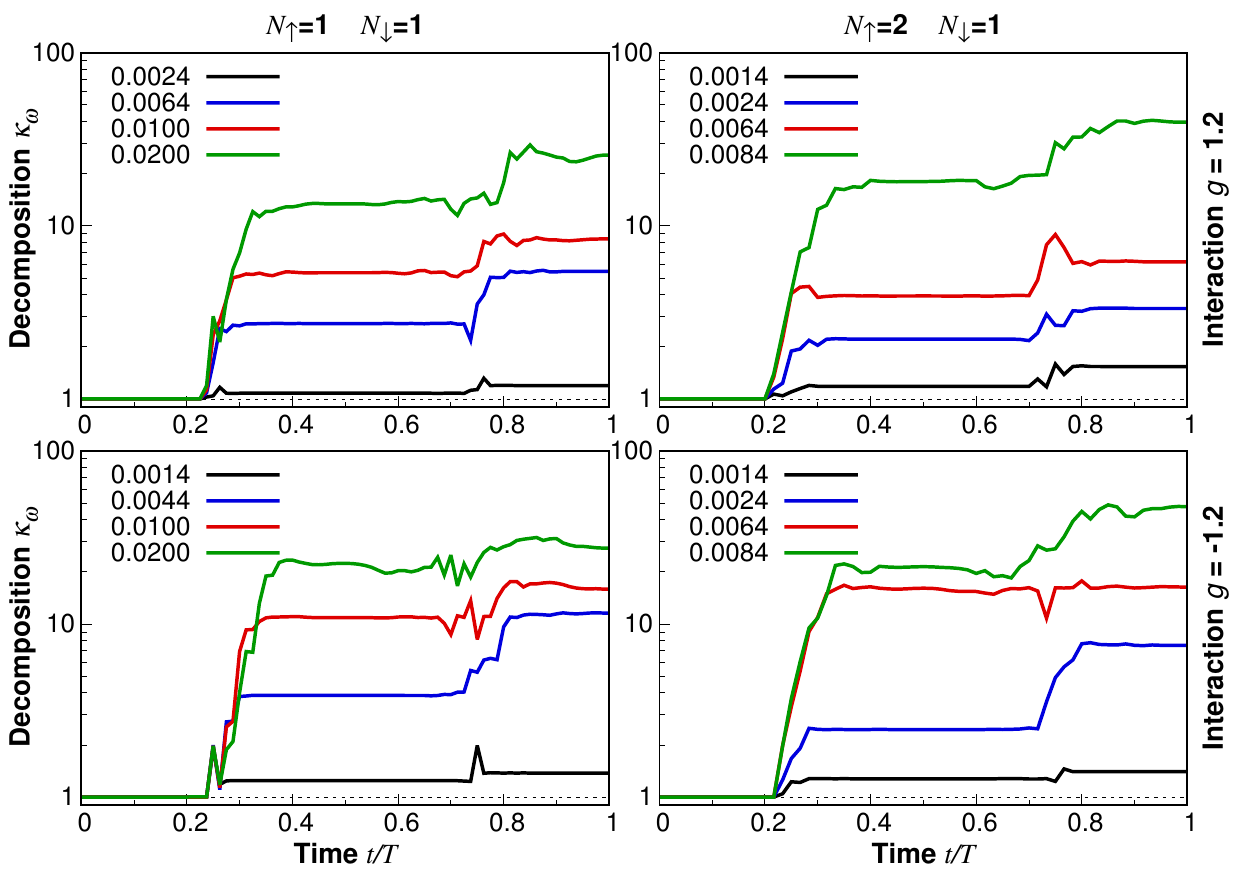}
\caption{Time evolution of the decomposition number $\kappa_\omega(t)$ for the balanced system of $N_\uparrow=N_\downarrow=1$ particles (left) and the imbalanced system of $N_\uparrow=2$ and $N_\downarrow=1$ particles (right). Depending on the character of mutual interactions (repulsive and attractive forces on top and bottom row, respectively) and frequency of the external driving different number of many-body eigenstates become occupied during the dynamics. Rapid changes of $\kappa_\omega$ occur close to the moments when quasi-degeneracies of the many-body spectrum are forced by moving barrier.} 
\label{Fig10}
\end{figure} 
To get better understanding of mentioned excitations to higher many-body states, let us also consider some generalisation of the concept of the temporal ground-state fidelity ${\cal F}_\omega(t)$ to fidelities with other many-body eigenstates of the system $|\mathtt{E}_k\rangle$. It can be done straightforwardly by introducing the excited-state fidelities ${\cal F}_\omega^{(k)}(t) = |\langle\psi(t)|\mathtt{E}_k(x_0(t))\rangle|^2$ (by definition, ${\cal F}_\omega^{(0)}(t)$ is equivalent to the ground-state fidelity ${\cal F}_\omega(t)$). With these quantities one can easily follow occupations of different many-body eigenstates during the evolution. However, when rapid drivings are considered, number of contributing many-body states becomes significant and precise analysis is difficult. Therefore, based on these fidelities, we introduce single quantity characterising global properties of the state's decomposition. Namely, we use the decomposition number $\kappa_\omega(t)$ defined as
\begin{equation}
\kappa_\omega(t) = \frac{1}{\sum_k \left[{\cal F}_\omega^{(k)}(t)\right]^2}.
\end{equation} 
The quantity has a very natural phenomenological interpretation. If the state is decomposed equally among $K$ different many-body states then $\kappa_\omega$ is equal to $K$. Therefore, $\kappa_\omega$ can be understood as an effective number of many-body states contributing to the many-body state. In Fig.~\ref{Fig10} we show the time-evolution of the decomposition number $\kappa_\omega(t)$ in the case of the archetypal balanced and imbalanced system (two and three particles) for two different interaction strength $g$ and different driving frequencies $\omega$. It is clear that along with increasing frequencies (more diabatic processes) larger number of temporal eigenstates become occupied. Note also that rapid changes of $\kappa_\omega$ occur at moments when the barrier's position leads to the quasi-degeneracy of the many-body spectra. These two facts are in full agreement with results for the ground-state fidelity ${\cal F}_\omega(t)$ (for example, compare with Fig.~\ref{Fig2}). In fact, the ground-state fidelity defines the upper bound for the decomposition number, $\kappa_\omega(t)\leq [{\cal F}_\omega(t)]^{-2}$.

\section{Conclusions}
In this work, we studied the dynamical resistivity of strongly correlated few-fermion systems to the driving by an external potential barrier. Assuming that the system is initially prepared in its interacting many-body ground state and by solving exactly the time-dependent many-body Schr\"odinger equation we examine how the system drives away from the temporal many-body ground state. It turned out that the dynamics of the system is always regular up to the instant when a quasi-degeneracy of the many-body spectrum is achieved. For later moments, due to a quite tangled relation between ballistic motion of the system, correlations induced by interactions, barrier position and speed, and corresponding changes of the many-body spectra the dynamics become highly unpredictable and very sensitive to different parameters. Importantly, moments in which the quasi-degeneracy of the spectrum is established highly depend on the number of particles and may occur more than once during the barrier transition. Moreover, we found that systems with an imbalanced number of particles have a non-vanishing energy gap between the ground and excited states for any interaction and any barrier's position. Therefore, imbalanced systems are much more resistant to the external driving. 

Our findings presented in this work may have importance for suspected experimental accuracy of time-dependent protocols designed for coherent manipulations of strongly correlated few-body systems. It turns out that, even for such tiny systems of a well-defined number of particles, initially prepared almost perfectly in their isolated ground-state, small and almost adiabatic external driving may lead to unpredictable dynamics and move the system far away from the desired state. The clue for solving this ambiguity is mostly encoded in changes of the temporal many-body spectrum. 

\section{Acknowledgments}
This work was supported by the (Polish) National Science Center Grant No. 2016/22/E/ST2/00555. Numerical calculations were partially carried out in the Interdisciplinary Centre for Mathematical and Computational Modelling, University of Warsaw (ICM), under Computational Grant No. G75-6. 

\section{References}
\bibliographystyle{iopart-num}
\bibliography{biblio}

\end{document}